\documentclass[conference]{IEEEtran}
\IEEEoverridecommandlockouts

\usepackage[english]{babel}

\usepackage{cite}
\usepackage{amsmath,amssymb,amsfonts}
\usepackage{algorithmic}
\usepackage{graphicx}
\usepackage{textcomp}
\usepackage{xcolor}
\usepackage{booktabs}
\usepackage{hyperref}
\def\BibTeX{{\rm B\kern-.05em{\sc i\kern-.025em b}\kern-.08em
    T\kern-.1667em\lower.7ex\hbox{E}\kern-.125emX}}

\usepackage{etoolbox}
\makeatletter
\patchcmd{\@makecaption}
  {\scshape}
  {}
  {}
  {}
\makeatother

\addto\extrasenglish{%
  
}

\addto\extrasenglish{
  
}

\usepackage{pifont}
\newcommand{\cmark}{\ding{51}}%
\newcommand{\xmark}{\ding{55}}%

\usepackage{multirow}

\begin{document}
\title{Can Foundation Models Really Segment Tumors? A Benchmarking Odyssey in Lung CT Imaging}

\author{
\IEEEauthorblockN{
Elena Mulero Ayllón\IEEEauthorrefmark{1},
Massimiliano Mantegna\IEEEauthorrefmark{1},
Linlin Shen\IEEEauthorrefmark{2},
Paolo Soda\IEEEauthorrefmark{1}\IEEEauthorrefmark{3}, \\
Valerio Guarrasi\IEEEauthorrefmark{1}\IEEEauthorrefmark{5}
and Matteo Tortora\IEEEauthorrefmark{4}\IEEEauthorrefmark{5}
}

\IEEEauthorblockA{
\IEEEauthorrefmark{1} Unit of Computer Systems and Bioinformatics, Department of Engineering, \\
Università Campus Bio-Medico di Roma,
Rome, Italy \\
Email: \{e.muleroayllon, m.mantegna,
valerio.guarrasi, p.soda\}@unicampus.it}

\IEEEauthorblockA{
\IEEEauthorrefmark{2} 
College of Computer Science and Software Engineering, Shenzhen University,
Shenzhen, China}

\IEEEauthorblockA{
\IEEEauthorrefmark{3} 
Department of Diagnostics and Intervention, Radiation Physics, Biomedical Engineering, \\
Umeå University, Umeå, Sweden }

\IEEEauthorblockA{
\IEEEauthorrefmark{4}
Department of Naval, Electrical, Electronics and Telecommunications Engineering, University of Genoa, Genoa, Italy
\\
Email: matteo.tortora@unige.it}

\IEEEauthorblockA{\IEEEauthorrefmark{5}
Contributed equally to this work}
}

\maketitle

\begin{abstract}
Accurate lung tumor segmentation is crucial for improving diagnosis, treatment planning, and patient outcomes in oncology. However, the complexity of tumor morphology, size, and location poses significant challenges for automated segmentation. This study presents a comprehensive benchmarking analysis of deep learning-based segmentation models, comparing traditional architectures such as U-Net and DeepLabV3, self-configuring models like nnUNet, and foundation models like MedSAM, and MedSAM~2. Evaluating performance across two lung tumor segmentation datasets, we assess segmentation accuracy and computational efficiency under various learning paradigms, including few-shot learning and fine-tuning. The results reveal that while traditional models struggle with tumor delineation, foundation models, particularly MedSAM~2, outperform them in both accuracy and computational efficiency. These findings underscore the potential of foundation models for lung tumor segmentation, highlighting their applicability in improving clinical workflows and patient outcomes.
\end{abstract}


\section{Introduction}
Lung cancer remains one of the most prevalent and deadly cancers worldwide, with early diagnosis playing a crucial role in improving patient outcomes~\cite{siegel2023cancer}. 
Computed tomography (CT) is the primary imaging modality for lung tumor detection and monitoring, offering high-resolution insights into tumor morphology~\cite{gulati2020lung}. 
However, manual segmentation of lung tumors is time-consuming and requires expert radiologists, often leading to inter-observer variability and inconsistencies in delineation~\cite{armato2011lung}. 
Consequently, automated lung tumor segmentation models are crucial for enhancing diagnostic efficiency and reproducibility in clinical workflows.

Deep learning-based methods have become increasingly prominent in the medical domain due to their ability to extract complex representations from heterogeneous data sources and support diverse clinical tasks, including treatment planning, outcome prediction, and disease characterization~\cite{tortora2021deep,nibid2023deep,liu2021exploring,furia2023exploring}.
Recently, foundation models have emerged as a promising paradigm, demonstrating strong generalization capabilities across multiple segmentation tasks without extensive task-specific retraining~\cite{bommasani2021opportunities}. 
These models leverage large-scale pretraining and transfer learning to adapt to new domains, making them particularly appealing for medical imaging applications. 
Notable examples include the Segment Anything Model (SAM), along with various medical imaging adaptations such as MedSAM~\cite{ma2024segment} and Medical SAM 2~\cite{zhu2024medical}, also referred to as MedSAM~2, which build upon SAM's framework~\cite{kirillov2023segment} and refine its performance in segmenting anatomical structures.

Despite their versatility, foundation models may struggle with zero-shot segmentation, where they are applied to tasks beyond their training distribution~\cite{shi2023generalist,dong2024efficient}. 
This challenge is particularly pronounced in lung tumor segmentation, where tumor heterogeneity, varying lesion sizes, and different growth patterns across cancer stages introduce complexities that general-purpose segmentation models may not fully capture~\cite{huang2024}. 
Furthermore, while traditional deep learning models such as U-Net~\cite{ronneberger2015u}, nnUNet~\cite{isensee2021nnu} and DeepLabV3~\cite{chen2017rethinking} have demonstrated strong performance in medical image segmentation, their effectiveness relative to foundation models remains an open question, particularly under different learning paradigms such as zero-shot, few-shot, and fine-tuning.

In this study, we present a comprehensive benchmarking analysis of state-of-the-art segmentation models, including traditional deep learning architectures (i.e., DeepLabV3, U-Net, nnUNet) and foundation models for medical imaging (i.e., MedSAM and MedSAM~2). 
Our contributions are as follows:
\begin{itemize}
    \item A comparative evaluation of segmentation models under few-shot and fine-tuning settings;
    \item Performance assessment across two different lung tumor segmentation datasets;
    \item Analysis of computational efficiency, examining the trade-offs between segmentation accuracy and computational cost;
    \item In-depth exploration of training strategies and prompting scenarios within the MedSAM~2 framework.
\end{itemize}

The remainder of this paper is structured as follows:~\autoref{sc:methods} describes the experimental setup, covering the segmentation models, training strategies, and evaluation metrics.~\autoref{sc:materials} describes the materials used in this study, including datasets and pre-processing steps.~\autoref{sc:results} presents the benchmarking results and comparative analysis of different models. 
Finally,~\autoref{sc:conclusion} discusses the key findings, limitations, and implications of the results, concluding the study with potential directions for future research.

\section{Methods and Experimental Setups}
\label{sc:methods}

To benchmark segmentation performance, we designed a comprehensive experimental framework encompassing model selection, evaluation strategy, and implementation details. We start by presenting the segmentation models selected for their relevance and architectural diversity. We then outline our experimental setup, including dataset characteristics, training configurations, and testing protocols. To evaluate performance, we define metrics that capture both segmentation accuracy and computational efficiency. Finally, we describe our training procedures, detailing hyperparameter choices, optimization strategies, and the computational resources employed.

\subsection{Benchmarking Models}
We selected a diverse set of segmentation models to serve as benchmarks, each representing different architectural paradigms and learning strategies. The models included in this study are:

\paragraph{DeepLabV3~\cite{chen2017rethinking}}
it is a deep convolutional neural network designed for semantic segmentation,
exhibiting strong performance across standard benchmarks.
In the medical domain, it has been adapted for several segmentation tasks~\cite{azad2022transdeeplab,wang2021medical,polat2022modified}, though it requires task-specific training to achieve optimal results.

\paragraph{U-Net~\cite{ronneberger2015u}} 
it is a fully convolutional network designed for biomedical image segmentation. 
Its encoder-decoder architecture with skip connections enables precise localization by combining spatial and contextual information.
It performs well even with limited training data, leveraging data augmentation, and has demonstrated strong results in domain-specific benchmarks.

\paragraph{nnUNet~\cite{isensee2021nnu}}
it is a self-configuring 
framework for biomedical image segmentation that automates the entire 
pipeline, including preprocessing, network architecture design, training, and post-processing. 
It systematically adapts to new datasets by leveraging a combination of fixed, rule-based, and empirical parameters. 
It has consistently outperformed task-specific methods across a wide range of benchmarks, establishing itself as a strong reference in the field.

\paragraph{MedSAM~\cite{ma2024segment}} it 
is a medical imaging adaptation of the Segment Anything Model (SAM)~\cite{kirillov2023segment}, incorporating domain-specific training to enhance anatomical structure segmentation in CT and MRI.
Leveraging large-scale pretraining and fine-tuning, it outperforms general-purpose models in organ and lesion segmentation tasks.

\paragraph{MedSAM~2~\cite{zhu2024medical}} 
it extends SAM2~\cite{ravi2024sam} by reframing medical segmentation as a video object tracking task.
It introduces a self-sorting memory bank to dynamically select relevant embeddings, enhancing performance on both 2D and 3D data. The model supports one-prompt segmentation and has demonstrated state-of-the-art results across diverse medical datasets.

\subsection{Experimental Setups}
To assess the performance of the segmentation models introduced in the previous section, we designed tailored training strategies and experimental configurations. Specifically, two distinct experiments were conducted to evaluate model effectiveness under varying conditions, reflecting both few-shot and fine-tuning scenarios.

In the first experiment, we aimed to assess the segmentation performance of the models under standard training procedures. 
For this purpose, we trained DeepLabV3 from scratch using a ResNet-101 backbone.
Similarly, the U-Net model was trained from scratch, with its architecture adapted following the approach described in~\cite{primakov2022automated}. 
For the nnUNet model, we adhered to its default training pipeline without modifications, exploring its three standard configurations: 2D, 3D low-resolution, and 3D full-resolution. 
In contrast, both MedSAM and MedSAM~2 were fine-tuned using their respective pre-trained weights. Specifically, for MedSAM~2, fine-tuning was performed using 50\% of the available training set.

Since this study also aims to investigate the applicability of foundation models in real-world scenarios, we conducted further experimental analysis specifically on MedSAM~2, the most recent foundation model in this domain.
In the second experiment, we focused exclusively on MedSAM~2 to investigate the impact of training data availability on model performance. 
Specifically, we conducted multiple training sessions using different fractions of the training set (25\%, 50\%, and 75\%). 
This approach enabled a thorough evaluation of the model’s robustness and adaptability to varying amounts of training data. 
MedSAM~2 was selected for this experiment as it demonstrated superior performance in preliminary evaluations, making it the most suitable candidate for analyzing the effect of training data availability. 
Additionally, since MedSAM~2 supports two prompting strategies—bounding box-based and click-based inputs—we trained the model using both configurations to analyze their influence on segmentation performance.

All experiments were conducted on two distinct lung tumor datasets, as described in~\autoref{sc:materials}. 
A summary of the training strategies employed for each model is provided in~\autoref{tab:exp_setup}, where an \xmark~symbol denotes the absence of a specific capability (e.g., zero-shot or prompt-based inference), and a \cmark~indicates its presence.

\begin{table}
\centering
\caption{Overview of the experimental setup for benchmarking models, indicating zero-shot or prompt-based inference capabilities and training strategies}
\label{tab:exp_setup}
\begin{tabular}{lccc}
\hline
\multicolumn{1}{c}{\textbf{Models}}     & \textbf{Zero-Shot} & \textbf{Prompt} & \textbf{Training} \\ \hline
DeepLabV3         & \xmark     & \xmark     & Scratch     \\
U-Net             & \xmark     & \xmark     & Scratch     \\
nnUNet 2d         & \xmark     & \xmark     & Scratch     \\
nnUNet 3d lowres  & \xmark     & \xmark     & Scratch     \\
nnUNet 3d fullres & \xmark     & \xmark     & Scratch     \\
MedSAM            & \cmark     & \cmark     & Fine-tune     \\
MedSAM~2          & \cmark     & \cmark     & Fine-tune     \\ \hline
\end{tabular}
\end{table} 

\subsection{Evaluation Metrics}
The performance of the segmentation models was evaluated using two widely recognized metrics: the Intersection over Union (IoU)~\cite{rezatofighi2019generalized} and the Dice Similarity Coefficient (Dice Score)~\cite{bertels2019optimizing}. 
Both metrics are commonly used in medical image segmentation tasks and provide complementary insights into the accuracy of the predicted segmentation masks relative to the ground truth. 
In the following equations, \(A\) represents the predicted segmentation, \(B\) represents the ground truth, and \(|A \cap B|\) is the area of overlap between the predicted and true regions, while \(|A \cup B|\) is the total area covered by either the predicted or the ground truth region:

\begin{itemize}
    \item \textbf{IoU:} it quantifies the overlap between the predicted segmentation and the ground truth~\cite{rezatofighi2019generalized}. 
    It is calculated as the ratio of the intersection of the predicted and ground truth regions to the union of those regions.
    \begin{equation} 
    \begin{split}
    IoU = \frac{|A \cap B|}{|A \cup B|}
    \end{split}
    \end{equation}
    
    \item \textbf{Dice Score:} it measures the similarity between the predicted and ground truth regions~\cite{bertels2019optimizing}. 
    It is calculated as twice the intersection of the predicted and ground truth regions divided by the sum of their areas.
    \begin{equation} 
    \begin{split}
    Dice\ Score = \frac{2|A \cap B|}{|A| + |B|}
    \end{split}
    \end{equation}
\end{itemize} 

\subsection{Training Details}

All models were trained, leveraging the official implementations provided by their respective authors, unless otherwise specified. 
The experiments were conducted on high-performance computing resources, utilizing different GPU architectures depending on the model requirements.
All models, except for nnUNet and MedSAM~2 trained on the Task06 dataset, were trained using an NVIDIA A100 GPU. 
The nnUNet model was trained on an NVIDIA T4, while the MedSAM~2 model trained on the Task06 dataset was executed on an NVIDIA A40 GPU.

For the training of nnUNet and MedSAM, all default hyperparameters were used without modifications, ensuring consistency with the original implementations.
The DeepLabV3 and U-Net models were trained for 300 epochs with a learning rate of 0.0001. 
Meanwhile, the MedSAM~2 model was trained for 1000 epochs, maintaining the default parameters for all other training configurations.

These training conditions were selected to ensure a fair and reproducible evaluation of each segmentation approach.

\section{Materials}
\label{sc:materials}

\subsection{Datasets}

The NSCLC-Radiomics dataset~\cite{aerts2019data}, also referred to as Lung1, consists of CT scans from 422 patients diagnosed with non-small cell lung cancer (NSCLC). 
Each scan includes a manual delineation of the 3D gross tumor volume. 
Due to inability to extract lung masks for some cases, we used a subset of 304 patients for our analysis. 
This dataset was split into a training set (246 patients) and a test set (58 patients).

The Task06 dataset from the Medical Segmentation Decathlon~\cite{antonelli2022medical} is a collection of 63 CT scans from patients diagnosed with NSCLC provided with delineations of small tumor volumes within the lungs. 
For our experiments, we split it into a training set (51 patients) and a test set (12 patients).

In both datasets, the splits were fixed across all experiments, preventing data leakage and enhancing reproducibility to ensure consistency and comparability.

\subsection{Pre-processing}

Since none of the datasets provide pre-existing lung masks, we first extracted lung masks directly from the CT images using the method proposed in~\cite{hofmanninger2020automatic}. 
The lung masks and tumor mask were then summed together, with each mask assigned a distinct pixel value to differentiate them. The resulting image is a single-channel representation containing all masks, where each mask corresponds to a unique intensity value.

To ensure consistency across all datasets, several pre-processing steps were applied. 
First, Hounsfield Unit conversion was performed, which maps CT intensity values to a standardized scale representing tissue densities. 
This conversion facilitates better contrast between different anatomical structures. 
Next, pixel spacing was resampled to (1, 1, 3) mm for all images to standardize voxel dimensions and maintain spatial consistency across datasets.

Additionally, image intensity values were clipped to the range [-1000, 1000] to suppress outlier values. 
Finally, normalization was applied to scale pixel values  to the range [0, 1], improving the stability of the models.

\section{Results}
\label{sc:results}

To compare the performance of the segmentation models, we present both quantitative and qualitative results. We first compare their overall performance on lung and tumor segmentation tasks, highlighting key findings from~\autoref{tab:exp_res} and ~\autoref{fig:inference-examples}. Next, we analyze their computational cost and efficiency, as illustrated in~\autoref{fig:computational-costs}. Finally, we investigate the impact of dataset size on the performance of the best-performing model, MedSAM~2, across different dataset splits in~\autoref{tab:exp_res_medsam_lung1} and~\autoref{tab:exp_res_medsam_task06}.

\subsection{Benchmarking model performance}
The results presented in~\autoref{tab:exp_res} highlight key observations regarding the performance of different models on lung and tumor segmentation tasks. 
When comparing all models, MedSAM~2 with bounding box prompts emerges as the top performer for tumor segmentation. 
Meanwhile, nnUNet achieves the best performance in lung segmentation and ranks as the second-best model for tumor segmentation. 
It is evident that all models achieve strong results in lung segmentation, a task that is relatively well-resolved in the literature due to the distinct and predictable anatomy of the lungs. 
However, DeepLabV3 and U-Net lag behind when it comes to tumor segmentation, performing the worst among the evaluated methods. 
Furthermore, tumor segmentation generally yields better results on Task06, where tumors are typically smaller and more centrally located, though this is not universally the case across all models. 

\begin{table*}
\centering
\caption{Performance comparison of benchmarking models on the Lung1 and Task06 datasets, evaluated using IoU and Dice score for lungs, tumor and average segmentation performance. Values in \textbf{bold} indicate the best performance, while \underline{underlined} values indicate the second-best.}
\label{tab:exp_res}
\resizebox{0.9\textwidth}{!}{
\begin{tabular}{lllllll|llllll}
\toprule 
  \multirow{3}{*}{\textbf{Methods}} & \multicolumn{6}{c|}{\textbf{Lung1}} & \multicolumn{6}{c}{\textbf{Task06 Lungs}}                      \\
           & \multicolumn{3}{c}{\textbf{IoU \(\uparrow\)} } & \multicolumn{3}{c|}{\textbf{Dice \(\uparrow\)}} & \multicolumn{3}{c}{\textbf{IoU \(\uparrow\)}} & \multicolumn{3}{c}{\textbf{Dice \(\uparrow\)}} \\
           & \textbf{Lungs}   & \textbf{Tumor}  & \textbf{Avg.}  & \textbf{Lungs}  & \textbf{Tumor}  & \textbf{Avg.}  & \textbf{Lungs}   & \textbf{Tumor}  & \textbf{Avg.}  & \textbf{Lungs}  & \textbf{Tumor}  & \textbf{Avg.}  \\
           \midrule

DeepLabV3             & 0.8763    & 0.0409    & 0.6970    & 0.9116    & 0.0532    & 0.8001    & 0.7021    & 0.0060    & 0.6016    & 0.7242    & 0.0087    & 0.6138  \\
U-Net                 & 0.8377    & 0.0430    & 0.8383    & 0.8832    & 0.0530    & 0.8938    & 0.6512    & 0.0131    & 0.6590    & 0.6874    & 0.0179    & 0.6995  \\
nnUnet 2d             & \textbf{0.9700}    & \underline{0.8442}    & \textbf{0.9281}    & \textbf{0.9844}    & \underline{0.9039}    & \textbf{0.9576}    & \textbf{0.9822}    & \underline{0.8023}    & \underline{0.9222}    & \textbf{0.9910}    & \underline{0.8736}    & \underline{0.9519}  \\
nnUnet 3d lowres      & \underline{0.9350}    & 0.6247    & 0.8316    & \underline{0.9619}    & 0.7386    & 0.8874    & \underline{0.9803}    & 0.7765    & 0.9123    & \underline{0.9900}    & 0.8650    & 0.9483    \\
nnUnet 3d fullres     & 0.9320    & 0.5912    & 0.8183    & 0.9601    & 0.7023    & 0.8742    & 0.9746    & 0.7515    & 0.9002    & 0.9871    & 0.8487    & 0.9409   \\
MedSAM                & 0.8648    & 0.5315    & 0.8236    & 0.9146    & 0.6441    & 0.8814    & 0.9228    & 0.6095    & 0.9018    & 0.9537    & 0.7230    & 0.9384  \\
MedSAM~2 Point & 0.7575    & 0.7349    & 0.7499    & 0.8208    & 0.7974    & 0.8130    & 0.8818    & 0.7770    & 0.8469    & 0.9053    & 0.7974    & 0.8693  \\
MedSAM~2 BBox  & 0.8857    & \textbf{0.8612}    & \underline{0.8775}    & 0.9342    & \textbf{0.9091}    & \underline{0.9258}    & 0.9712    & \textbf{0.8536}    & \textbf{0.9321}    & 0.9980    & \textbf{0.8770}    & \textbf{0.9577}  \\
\bottomrule
\end{tabular}
}
\end{table*}

\begin{figure*}    \includegraphics[width=0.85\textwidth]{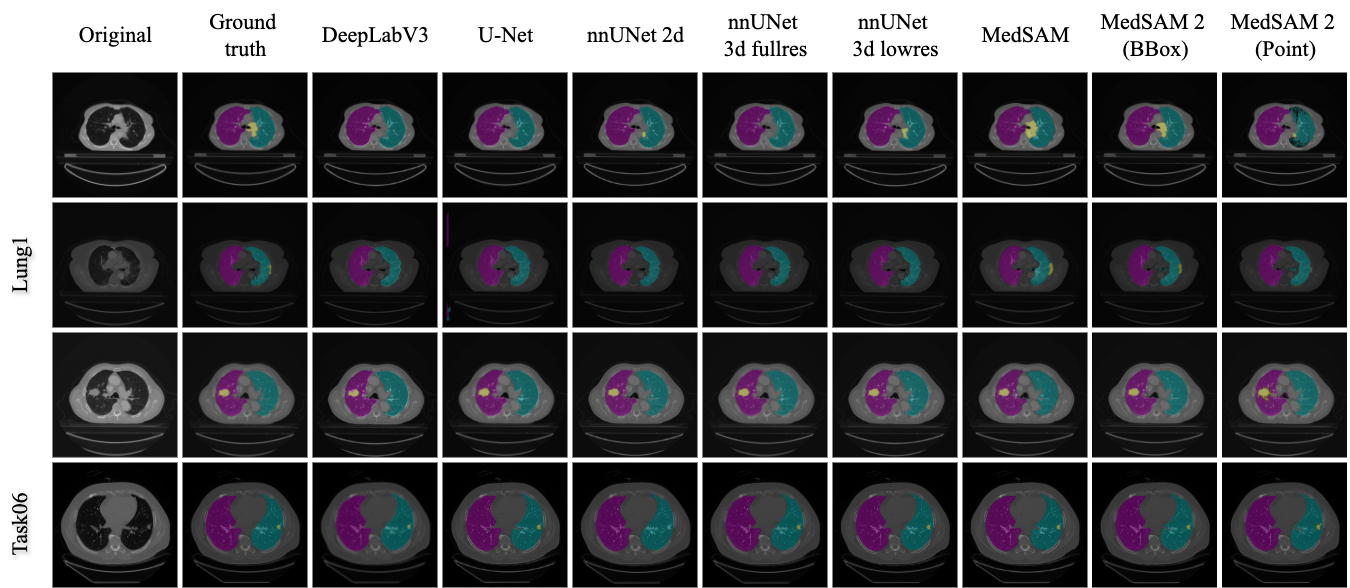}
    \centering  
    \caption{Qualitative comparison of segmentation results across different models. The first column presents the original CT scan images, followed by the ground truth segmentations of left and right lungs and tumor mass. The remaining columns showcase the predictions generated by the benchmarking models.}
    \label{fig:inference-examples}
\end{figure*}

The qualitative results presented in~\autoref{fig:inference-examples} further illustrate the segmentation performance of the evaluated models. 
Irregular and non-centered tumor masses are not accurately segmented by models such as nnUNet or MedSAM, while DeepLabV3 and U-Net fail to detect them altogether. 
Notably, only MedSAM~2, when using bounding box prompts, achieves accurate segmentation in these challenging cases, as seen in the first two example images. 
In contrast, when tumor masses are well-defined and centrally located within the lungs, most models are capable of detecting them effectively, as demonstrated in the last two example images. 
Regarding lung segmentation, all models consistently achieve high accuracy, with the exception of MedSAM~2, which may occasionally fail when using point-based prompts.

\begin{figure}
\includegraphics[width=\columnwidth]{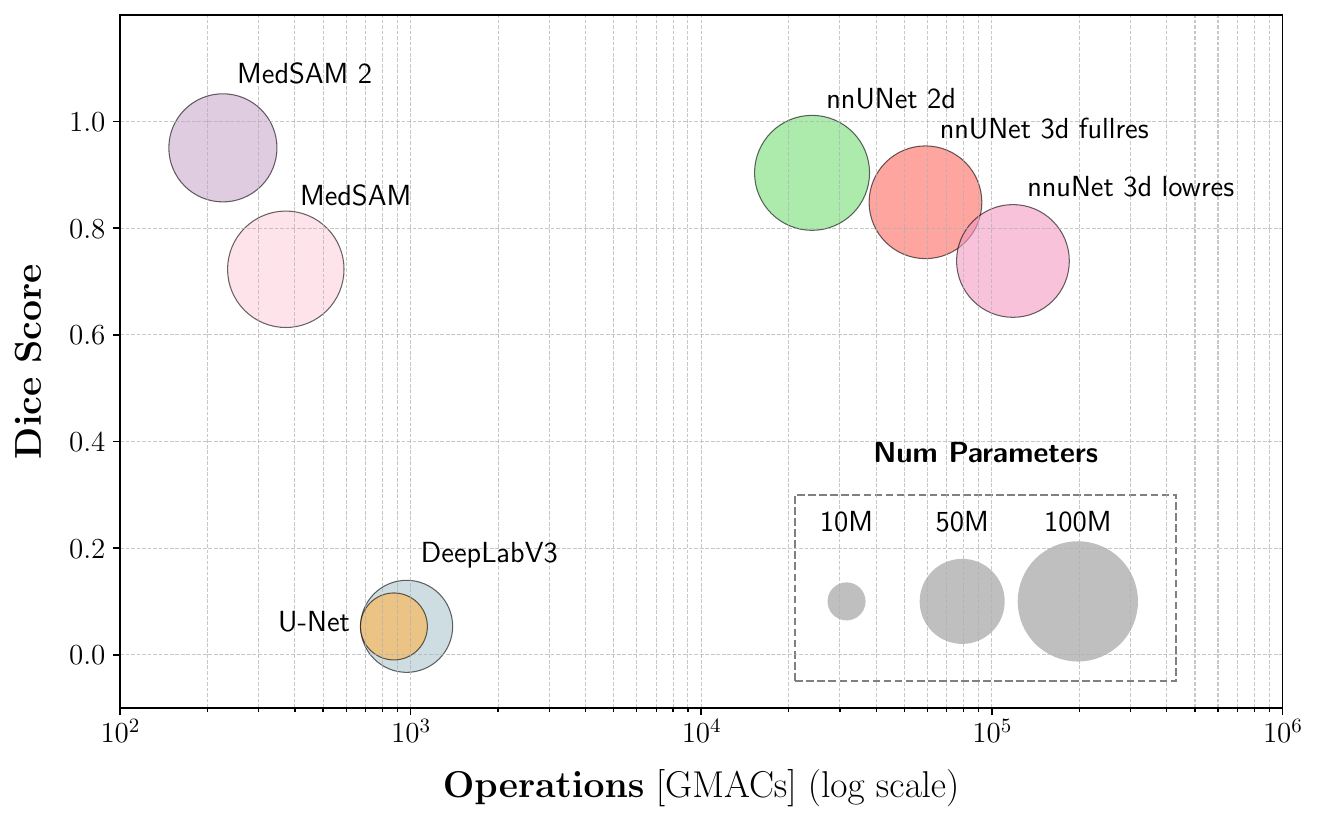}
\centering
\caption{Comparison of segmentation models in terms of Dice Score (y-axis) and computational cost measured in GMACs (x-axis, log scale). 
The size of each bubble is proportional to the number of model parameters, as illustrated by the gray reference bubbles in the bottom right corner, corresponding to 10M, 50M, and 100M parameters.}
\label{fig:computational-costs}
\end{figure}

~\autoref{fig:computational-costs} shows a clear trade-off between computational cost and segmentation performance. 
MedSAM~2 achieves the highest Dice score with relatively low computational cost (aproximately 226 GMACs~\cite{sun2024efficient}), making it the most efficient model.
nnUNet models exhibit strong performance but at a significantly higher computational cost. 
nnUNet 2D reaches a slightly lower Dice score but demands 24,062 GMACs, while the 3D full-resolution and low-resolution variants require even more resources (59,097 and 118,194 GMACs, respectively) for lower Dice scores.
Traditional models like DeepLabV3 and U-Net perform poorly despite their lower computational costs, indicating their limitations for lung tumor segmentation.
Overall, MedSAM~2 provides the best balance of accuracy and efficiency, while nnUNet models achieve high performance at a much higher computational cost.

\subsection{MedSAM~2 analysis}

Since MedSAM~2 is the most effective model for lung tumor segmentation, both qualitatively and quantitatively, as demonstrated in the previous results, its performance across different dataset splits warrants further analysis. 
The results presented in~\autoref{tab:exp_res_medsam_lung1} (Lung1) and~\autoref{tab:exp_res_medsam_task06} (Task06) reveal several key trends. 
For Lung1, both bounding box and point prompting strategies show improved performance in all segmentation tasks as the percentage of the training dataset increases. 
This improvement continues until the dataset reaches a point, typically between 50\% and 75\%, where overfitting may occur. 
Overfitting in such cases may stem from the model becoming too specialized to the training data, unable to generalize well to unseen data due to the limited diversity in smaller datasets. 
On Task06, a similar pattern is observed, where performance improves as the dataset size increases, with overfitting generally occurring around the 75\% split in many cases. 
These results suggest that a relatively small number of samples is sufficient for effective model training, pointing to the efficiency of MedSAM~2 in learning from a limited dataset. 
Interestingly, tumor segmentation performance using bounding box prompts in Task06 is better when using the model’s original weights, without fine-tuning, which may reflect the robustness of the pretrained model to the task, particularly when the tumors are well-defined.
Moreover, across both datasets and all segmentation tasks, bounding box prompts consistently yield superior performance compared to point prompts. 
This may be attributed to the more comprehensive spatial information provided by bounding boxes, which offers a more direct and structured way to guide the model, leading to more accurate segmentation outcomes.

\begin{table*}
\centering
\caption{Performance comparison of MedSAM~2 on different percentages of Lung1 training dataset, evaluated using IoU and Dice score for lungs, tumor and average segmentation performance. Values in \textbf{bold} indicate the best performance, while \underline{underlined} values indicate the second-best.}
\label{tab:exp_res_medsam_lung1}
\resizebox{0.9\textwidth}{!}{
\begin{tabular}{lllllll|llllll}
\toprule
 \multirow{3}{*}{\textbf{Methods}}         & \multicolumn{6}{c|}{\textbf{Bounding Box}}                      & \multicolumn{6}{c}{\textbf{Point}}                      \\
           & \multicolumn{3}{c}{\textbf{IoU \(\uparrow\)}} & \multicolumn{3}{c|}{\textbf{Dice \(\uparrow\)}} & \multicolumn{3}{c}{\textbf{IoU \(\uparrow\)}} & \multicolumn{3}{c}{\textbf{Dice \(\uparrow\)}} \\
           & \textbf{Lungs}   & \textbf{Tumor}  & \textbf{Avg.}  & \textbf{Lungs}  & \textbf{Tumor}  & \textbf{Avg.}  & \textbf{Lungs}   & \textbf{Tumor}  & \textbf{Avg.}  & \textbf{Lungs}  & \textbf{Tumor}  & \textbf{Avg.}  \\
           \midrule

0   & 0.8630    &    0.8610    & 0.8622    & 0.9103    & \underline{0.9092}    & 0.9099    & 0.5101    & 0.5009    & 0.5071    & 0.5870    & 0.5769    & 0.5837 \\
25  & 0.8761    & 0.8517    & 0.8680    & 0.9257    & 0.9006    & 0.9173    & 0.7646    & 0.7420    & 0.7571    & 0.8279    & 0.8045    & \underline{0.8201}  \\
50  & \textbf{0.8857}    & \textbf{0.8612}    & \textbf{0.8775}    & \textbf{0.9342}    & \textbf{0.9091}    & \textbf{0.9258}    & 0.7575    & 0.7349    & 0.7499    & 0.8208    & 0.7974    & 0.8130  \\
75  & \underline{0.8830}    & \underline{0.8588}    & \underline{0.8750}    & \underline{0.9319}    & 0.9070    & \underline{0.9236}    & \textbf{0.7893}    & \textbf{0.7661}    & \textbf{0.7816}    & \textbf{0.8505}    & \underline{0.8264}    & \textbf{0.8425}  \\
100 & 0.8285    & 0.8044    & 0.8205    & 0.8812    & 0.8563    & 0.8729    & \underline{0.7882}    & \underline{0.7654}    & \underline{0.7806}    & \underline{0.8504}    & \textbf{0.8268}    & \textbf{0.8425}  \\
\bottomrule
\end{tabular}
}
\end{table*}

\begin{table*}
\centering
\caption{Performance comparison of MedSAM~2 on different percentages of Task06 training dataset, evaluated using IoU and Dice score for lungs, tumor and average segmentation performance. Values in \textbf{bold} indicate the best performance, while \underline{underlined} values indicate the second-best.}
\label{tab:exp_res_medsam_task06}
\resizebox{0.9\textwidth}{!}{
\begin{tabular}{lllllll|llllll}
\toprule
 \multirow{3}{*}{\textbf{Methods}}         & \multicolumn{6}{c|}{\textbf{Bounding Box}}                      & \multicolumn{6}{c}{\textbf{Point}}                      \\
           & \multicolumn{3}{c}{\textbf{IoU \(\uparrow\)}} & \multicolumn{3}{c|}{\textbf{Dice \(\uparrow\)}} & \multicolumn{3}{c}{\textbf{IoU \(\uparrow\)}} & \multicolumn{3}{c}{\textbf{Dice \(\uparrow\)}} \\
           & \textbf{Lungs}   & \textbf{Tumor}  & \textbf{Avg.}  & \textbf{Lungs}  & \textbf{Tumor}  & \textbf{Avg.}  & \textbf{Lungs}   & \textbf{Tumor}  & \textbf{Avg.}  & \textbf{Lungs}  & \textbf{Tumor}  & \textbf{Avg.}  \\
           \midrule

0   & 0.9302    & \textbf{0.9217}    & 0.9274    & 0.9572    & \textbf{0.9508}    & 0.9550    & 0.7501    & 0.7116    & 0.7372    & 0.8012    & 0.7625    & 0.7883    \\
25  & 0.9683    & 0.8537    & 0.9301    & 0.9967    & 0.8780    & 0.9571    & 0.8652    & 0.7675    & 0.8327    & 0.8879    & 0.7881    & 0.8546  \\
50  & 0.9712    & 0.8536    & \underline{0.9321}    & 0.9980    & 0.8770    & 0.9577    & 0.8818    & 0.7770    & 0.8469     & 0.9053    & 0.7974    & 0.8693  \\
75  & \underline{0.9747}    & \underline{0.8578}    & \textbf{0.9357}    & \textbf{0.9989}    & \underline{0.8816}    & \textbf{0.9618}    & \textbf{0.8872}    & \textbf{0.7861}    & \textbf{0.8536}    & \textbf{0.9134}    & \underline{0.8088}    & \textbf{0.8785}  \\
100 & \textbf{0.9749}    & 0.8576    & \textbf{0.9357}    & \underline{0.9985}    & 0.8812    & \underline{0.9616}    & \underline{0.8858}    & \underline{0.7855}    & \underline{0.8524}    & \underline{0.9126}    & \textbf{0.8094}    & \underline{0.8782}  \\
\bottomrule
\end{tabular}
}
\end{table*}

\section{Conclusion}
\label{sc:conclusion}

In this study, we conducted a comprehensive benchmarking analysis of various segmentation models for lung tumor segmentation, comparing traditional deep learning architectures such as U-Net, DeepLabV3, and nnUNet with foundation models like MedSAM, and MedSAM~2. 
Our evaluation encompassed different learning paradigms, including few-shot learning and fine-tuning, across two lung tumor segmentation datasets. 
Through rigorous experimentation, we analyzed the trade-offs between segmentation accuracy and computational efficiency, identifying MedSAM~2 as the most effective model in terms of segmentation performance and computational cost.

Our findings highlight several key observations. 
While all models performed well in lung segmentation, tumor segmentation remained significantly more challenging due to variations in tumor morphology, location, and size. Traditional deep learning models such as U-Net and DeepLabV3 exhibited suboptimal performance, struggling with accurate tumor delineation, whereas foundation models, particularly MedSAM~2 with bounding box prompts, demonstrated superior performance. 
Notably, MedSAM~2 achieved the best segmentation results while maintaining a relatively low computational cost, making it a promising candidate for real-world medical imaging applications.

Despite these promising results, several limitations must be acknowledged. 
First, while MedSAM~2 demonstrated strong segmentation capabilities, its reliance on user-defined prompts introduces challenges in clinical workflows, where precise and consistent prompt annotations may not always be readily available. 
Additionally, our study focused on two specific datasets, and the generalizability of our findings to other medical imaging modalities or broader patient populations remains an open question. 
Furthermore, our experiments primarily examined MedSAM~2 under controlled conditions (i.e., using precise bounding boxes) and its real-world deployment in clinical settings may require further optimization and validation.

Future research should explore strategies to automate prompt generation, reducing reliance on manual annotations and improving the usability of foundation models in clinical environments. Extending this benchmarking study to larger, more diverse datasets with different anatomical structures or organs and integrating federated learning techniques could provide deeper insights into the robustness and scalability of foundation models for medical image segmentation.

In conclusion, our study underscores the potential of foundation models, particularly MedSAM~2, in advancing lung tumor segmentation. While challenges remain, continued research and methodological improvements could pave the way for their integration into clinical workflows, ultimately enhancing diagnostic accuracy and patient outcomes.

\section*{Acknowledgments}
Massimiliano Mantegna is a PhD student enrolled in the National PhD program in Artificial Intelligence, XXXVIII cycle, course on Health and life sciences, organized by Università Campus Bio-Medico di Roma. 
This work was partially supported by:
i) the Italian Ministry of Foreign Affairs and International Cooperation, grant
number PGR01156, ii) PNRR MUR project PE0000013FAIR, iii)  Università Campus Bio-Medico di Roma within the project “AI-powered Digital Twin for next-generation lung cancEr cAre (IDEA).
Resources are provided by the National Academic Infrastructure for Supercomputing in Sweden (NAISS) and the Swedish National Infrastructure for Computing (SNIC) at Alvis @ C3SE, partially funded by the Swedish Research Council through grant agreement no. 2022-06725 and no. 2018-05973.

\bibliographystyle{IEEEtran}
\bibliography{IEEEabrv, references.bib}

\end{document}